\documentclass[paper,11pt]{JHEP}
\usepackage{epsfig}

\renewcommand{\a}{\alpha}

\newcommand{\m}{\mu}

\newcommand{\rmd}{{\rm d}}

\def\dslash{\not{\hbox{\kern-2pt $\partial$}}}
\def\eslash{\not{\hbox{\kern-2pt $\epsilon$}}}
\def\Dslash{\not{\hbox{\kern-4pt $D$}}}
\def\Aslash{\not{\hbox{\kern-4pt $A$}}}
\def\Qslash{\not{\hbox{\kern-4pt $Q$}}}
\def\Wslash{\not{\hbox{\kern-4pt $W$}}}
\def\pslash{\not{\hbox{\kern-2.3pt $p$}}}
\def\kslash{\not{\hbox{\kern-2.3pt $k$}}}
\def\qslash{\not{\hbox{\kern-2.3pt $q$}}}

\def\be{\begin{equation}}
\def\ee{\end{equation}}
\def\ba{\begin{array}}
\def\ea{\end{array}}
\def\bea{\begin{eqnarray}}
\def\eea{\end{eqnarray}}

\def\Tr{{\rm Tr}}

\title{Stationary RG Flow and Thermodynamics of
Conformal Field Theories \thanks{Research
supported in part by the US Department of Energy
under grant DE--FG05--91ER40627 and by the Greek National Fellowships
Foundation.}}

\author{ Anastasios C. Petkou\\
Department of Physics, Theoretical Physics\\
University of Kaiserslautern, Postfach 3049\\
Kaiserslautern 67653, Germany\\
E-mail: \email{petkou@physik.uni-kl.de}}

\author{George Siopsis\\
Department of Physics and Astronomy, \\
The University of Tennessee, \\
Knoxville, TN 37996--1200, U.S.A.\\
E-Mail: \email{gsiopsis@utk.edu}}
\date{\today}
\preprint{UTHET--99--0501, THES-TP 99/05}

\abstract{
We argue that one can understand the relationship between 
critical free-energy densities of theories connected via
non-perturbative renormalization group flows, in terms of the
the number of fields coupled to the class of different theories
which flow into the same stationary renormalization group
trajectory. As an explicit example, we study the free-energy density
of bosonic and fermionic theories 
possessing strongly coupled critical points in $D=3$. We construct a
stationary trajectory which interpolates between
the free massless theory of $N$ scalars and a class of interacting
theories including both bosons and fermions. The free-energy along
this trajectory remains constant, but the degrees of freedom coupled
to the underlying theories which flow into it are
different. The difference in the degrees of freedom 
underlying two distinct points of the stationary trajectory, 
provides a natural explanation for the rational relationship between the
critical free-energy densities of the free massless theory of $N$
scalars and the $O(N)$ vector model. }

\newpage

\begin{document}

\section{Introduction}

The recent progress in our understanding of the microscopic origin
of the thermodynamic properties of black holes has sparked a renewed
interest in the investigation of conformal field theories (CFTs) at finite
temperature.
There have been a number of
calculations \cite{Klebanov,Gubser,Fotopoulos,Mozo,Rey} regarding the
free-energy density of CFTs emerging 
from adS/CFT correspondence
\cite{Maldacena,Polyakov,Witten}. These are particularly interesting,
as the adS/CFT correspondence probes the strong-coupling regime of
${\cal{N}}=4$ SYM$_{4}$ for 
large-$N$. The free-energy obtained is found to be 3/4
times the free-energy density of the same theory at its free-field limit.
Such a result still lacks a proper understanding since it is related
to non-trivial $D$-dimensional 
conformal field theories, where $D>2$, which are not well understood.

It has also
been suggested in \cite{Appelquist}, following earlier work of
\cite{Cardy1,Fradkin,Sachdev1,Tassos1,Tassos2},  that the free-energy density
at criticality encodes important information for
the RG-flow between different  fixed points.\footnote{For a 
different approach see \cite{Jose}.} In particular, it has been
claimed in \cite{Appelquist} that studies of the free-energy density
yield constraints regarding the low-energy spectrum of theories undergoing 
symmetry-breaking transitions. 

To get more insight into the significance of the free-energy density
at criticality, we elaborate here on some remarkable results 
regarding the three-dimensional $O(N)$ vector 
model and its fermionic counterpart, the Gross-Neveu model.
These models have non-trivial critical points at large-$N$ which
are strongly coupled and represent a symmetry-breaking transition,
making  them important testing ground. A remarkable result
is then that the leading-$N$ 
free-energy density of the bosonic theory
at its non-trivial critical point is 4/5 times the free-energy
density of the theory of $N$ massless free scalars
\cite{Sachdev1}.  

Our purpose here
is to provide some physical insight into the origin of this simple
relationship between the two critical points. To this end,
we construct a theory of coupled bosonic and fermionic fields.
Then, on the space of couplings we identify a stationary trajectory which 
interpolates between the free massless theory of $N$ scalars and a
class of interactive critical theories with an $O(N/4)$ symmetry. The
free-energy along this 
trajectory does not change. Its value remains equal to the
free-energy of the theory of $N$ massless free scalars.
At one end of the trajectory only four $O(N/4)$ massless fields
contribute and therefore the underlying theory is that of $N$ massless
free scalars.
We then find a point on the
trajectory at which the underlying theory is that of five copies of
the $O(N/4)$ vector model. 
This result explains why the 
free-energy density at the non-trivial critical point of the $O(N)$
vector model is 4/5 times
the free-energy
density of the free theory of $N$ massless scalars.
Our method
is quite general and could in principle be used to studying other
theories which have strongly coupled critical points, such as
${\cal{N}}=4$ SYM$_{4}$ in the large-$N$ limit.

We organize our discussion as follows. In Section~\ref{sec1}, we
introduce the free-energy
density and briefly discuss its properties. In Section~\ref{sec2}, we
review the salient features of the 
$O(N)$ vector model as well as the fermionic Gross-Neveu model.
Section~\ref{sec3} contains our main calculation. We construct an
interactive field
theory in three dimensions, making use of the models in Section~\ref{sec2}.
We show that this theory possesses stationary trajectories along
which the free-energy does not change. We identify a point on one of
the trajectories whose free-energy is manifestly 5/4 times the free-energy
of the non-trivial critical point of the $O(N)$ vector model. 
Finally, in Section~\ref{sec4}, we present our conclusions.
   
\section{General Remarks on the Free-Energy Density}
\label{sec1}

Consider a Euclidean field theory in ${\bf{S^{1}\times
R^{d-1}}}$ geometry with periodic boundary conditions. Such a theory
is related to a $d$-dimensional 
statistical mechanical system in thermal
equilibrium, as the lattice spacing $a\rightarrow 0$. The
free-energy density (free-energy per unit volume), for the latter
system  can be written as
\cite{Fradkin} 
\begin{equation}
\label{ena}
f(L,\tilde{g},a)=f_{0}(\tilde{g},a)-\frac{1}{L^{d}}C(L,\tilde{g},a)
\end{equation}
where the temperature is $T=1/L$ and $\tilde{g}$ denotes collectively
the bare coupling constants. The term $f_{0}$ is  the free-energy
density at zero temperature, or equivalently the ``bulk'' free-energy
density. 

Under the basic assumption that in the limit $a\rightarrow 0$ the statistical
mechanical system 
is described by a renormalizable QFT, UV
renormalization of $f_{0}$ suffices to render Eq.~(\ref{ena})
finite. Furthermore, one can write 
\begin{equation}
C(L,\tilde{g},a)\equiv C_{R}(g_{R},\m L)
\end{equation}
where $g_{R}$ are dimensionless coupling constants and $\m$ is a mass
scale. The subscript $R$
denotes the renormalized quantities. It is important to stress that
the dimensionless quantity $l=\m L$ is {\it not} the standard RG
scale since $L^{-1}$ is a physical mass scale not related to the
UV cut-off. Nevertheless, as $C_R$ is an observable it should not
depend on $\m$ and should therefore satisfy the following RG equation
(we drop the $R$ subscript since we shall henceforth deal only
with renormalized physical quantities)
\be
\label{duo}
\left( l\frac{\partial}{\partial l}+\beta(g)\frac{\partial}{\partial
g}\right) C(g,l)= 0
\ee
\be
\beta(g)= \m\frac{\partial g(l)}{\partial \m}
\ee
Eq.(\ref{duo}) is solved \cite{Fradkin} by the introduction of the
inverse correlation length $\xi^{-1}(g,l)=M(g,l)$ which has dimensions
of mass and obeys the following RG equation
\be
\label{corlength}
\left( l\frac{\partial}{\partial l}+\beta(g)\frac{\partial}{\partial
g}\right) M(g,l)= 0
\ee
Then, from Eq.~(\ref{duo}) and Eq.(\ref{corlength}) we may write
\be
\label{cR}
C(g,l)\equiv C(ML)
\ee
Denoting  $t=ML$, then $C(t)$ is a RG-invariant quantity.

The critical points of the theory correspond to certain values $g_*$
of the coupling(s) such that 
\be
\beta(g_*)=0 \label{beta_f}
\ee
From Eq.~(\ref{corlength}), this  implies the existence of a critical
inverse correlation length $M_*= M(g_*)$ and therefore of a critical
value for $C(t_*)$, $t_* = M_* L$. It can then be seen from
Eq.~(\ref{duo}) and Eq.~(\ref{corlength}) that the possible values of
the critical correlation length - which would correspond to possible
critical points of the theory - are given by the stationary points of
$C(t)$. 

If Eq.~(\ref{beta_f}) has more than one
solutions, they must correspond to different critical points. The
introduction of the length scale $L$, being essential 
for the existence of more that one stationary points of $C(t)$,
provides the means of connecting the two theories flowing to the different
critical points. These two critical points may in general have
different massless degrees of freedom coupled into them and one might be
tempted to view them as UV and IR critical points of the {\it same}
theory. However, it is rather difficult to demonstrate the
existence of such a RG
flow as it is in principle non-perturbative. An explicit example of such
a case is provided by the $O(N)$ $\phi^4$ theory 
and the $O(N)$ Vector Model in $d=3$. The latter theory has a
non-trivial UV fixed point which has been argued to be the same as
the IR fixed point of the former theory. This
will be discussed in Section~\ref{sec2}.

Nevertheless, Eq.~(\ref{duo}) can still provide important insight towards
the understanding of the RG flow between critical
points having different massless degrees of freedom coupled into them.
The essential observation is that Eq.~(\ref{duo}) 
allows the construction of  
stationary flows connecting theories with different field
content. Indeed, had we been 
able to construct a function $C[t(t')]$ which is stationary for all
values of $t'$ in a certain interval, then, from Eq.~(\ref{beta_f}), this
would correspond to a line of critical points parametrized by
the new parameter $t'$. To all these critical points corresponds the
same value of 
$C$, which is essentially a constant, therefore they have the
same massless degrees of freedom coupled into them. In other words, $t'$ 
now parametrizes different 
underlying theories flowing to the same critical point. This way we
are able to deduce simple relations between the degrees of freedom
coupled at the critical points of different theories.

We mentioned above that the quantity $C(g,l)$ carries information for
all the degrees of 
freedom coupled to the theory at a certain RG-scale. The reason is
that  $C(g,l)$ is 
related to the energy momentum tensor.
Consider a system in a ``slab'' of length $L$ with
partition function
\begin{equation}
Z_{L}=\int {\cal{D}}\phi~e^{-S_{L}(\phi)}=e^{-F(L)}
\end{equation}
where $F(L)$ is the free-energy normalized so that
\be F(\infty)=0\ee
Under the
coordinate change,
\begin{equation}
x_{\mu}\rightarrow x_{\mu}+\alpha_{\mu}(x)\,,\,\,\,\,\,
\alpha_{\mu}(x)= \left[\; \left(1+\frac{\Delta
L}{L}\right)x_{0}, \quad \bar{x}\right]\,, \,\,\,\,
\bar{x}=(x_1,..,x_{d-1})
\end{equation}
we have
\begin{equation}
S_{L}\rightarrow S_{L}+\frac{\Delta L}{L}\int_{slab}\rmd^{d}x \,T_{00}(x)
\end{equation}
If the partition function is at a fixed point of the RG, it  
remains invariant
under such a change. This gives to leading order in $\Delta L$
\begin{equation}
Z_{L}=e^{-F(L)}=e^{-F(L)-\Delta L\frac{\partial
F}{\partial L}-\frac{\Delta L}{L}\int_{slab}
\rmd^{d}x\langle T_{00}(x)\rangle}
\end{equation}
which gives the change of  the free-energy in terms of the $T_{00}$-component
of the energy momentum tensor as \cite{Cardy1}
\begin{equation}
L\frac{\partial F}{\partial L} =-\int_{slab}\rmd^{d}x\langle T_{00}(x)\rangle
\end{equation}
Using Eq.~(\ref{ena}), we then obtain
\begin{equation}
\label{tria}
\langle T_{00}\rangle
=(1-d)\frac{C(g,l)}{L^{d}}+\frac{1}{L^{d}}\beta(g)
\frac{\partial}{\partial g}C(g,l)
\end{equation}
Relations such as Eq.~(\ref{tria}), acquire their full power in two
dimensions where
they lead to the important result that $C(g,l)$ at a critical point is
proportional to the central charge of the corresponding CFT.
  
\section{Critical Bosonic and Fermionic Theories in $D=3$}
\label{sec2}

In this Section, we review $O(N)$ bosonic and $U(N)$ fermionic models in the
large-$N$ limit.

\subsection{The $O(N)$ Vector Model}

The Lagrangian density for the $O(N)$ bosonic vector model is
\be
{\cal L} = {1\over 2} \partial_\mu \vec\phi\cdot \partial^\mu \vec\phi
+ V(|\vec\phi|)
\ee
where $\vec\phi(x)=\phi^a(x)$, $a=1,2,..,N$ is the $O(N)$-vector scalar
field, and $V(|\vec\phi|)$ is an interaction potential whose explicit form
is not important for our purposes. The partition function in the large-$N$
limit exhibits a universal behavior. This has been studied in the case when
\be
V(|\vec\phi|) = \lambda (\vec\phi\cdot \vec\phi)^2
\ee
and also when the vector field is constrained by
\be\vec\phi\cdot \vec\phi =1\ee
In the latter case, the Lagrangian density may be written in terms of
an auxiliary scalar field $\sigma$ as
\be
\label{eqlang}
{\cal L} = {1\over 2}\;\vec\phi \cdot (-\partial^{2}
+\sigma)\vec\phi + {N\over g_0} \sigma
\ee
This theory possesses a non-trivial critical point at each
order in a $1/N$ expansion, as the coupling constant
is kept fixed \cite{Zinn-Justin,Rosenstein}. This critical point is UV-stable,
corresponds to a 
non-trivial three-dimensional CFT and represents the
$O(N)\rightarrow O(N-1)$ symmetry
breaking transition. The critical theory is strongly coupled, since
the $1/N$ expansion corresponds to a resummation of an infinite number
of diagrams. 

In the spirit of the suggestion in \cite{Appelquist}, we can measure
the massless degrees of freedom 
coupled to a critical theory if we heat it up to a temperature $T = 1/L$
where we impose periodic boundary conditions along Euclidean time.
This is equivalent to  
placing the theory in a slab with one
finite dimension of length $L$ playing the role of inverse
temperature. 
Starting from the Lagrangian~(\ref{eqlang}) and using a saddle
point approximation, we obtain the free energy density
\be
f(L,g_0) = {N\over 2L} \;\sum_{n=-\infty}^\infty \int {d^2 \bar p\over
  (2\pi)^2} 
\ln (\bar p^2 + \omega_n^2 +\sigma_s) - {N\over g_0} \; \sigma_s
\ee
where $\bar{p}=(p_{1},p_{2})$ and $\sigma_s$ is the saddle-point value
of the auxiliary field $\sigma$ 
constrained by the gap equation
\be\label{eqgap}
{\partial f\over \partial\sigma_s} = {N\over 2L}
\;\sum_{n=-\infty}^\infty \int {d^2 \bar p\over (2\pi)^2} 
{1\over \bar p^2 + \omega_n^2 +\sigma_s} - {N\over g_0} = 0
\ee
The bare coupling constant $g_0$ is renormalized by the RGE
\be\label{eqRG}
{1\over g_R} = {1\over g_0} - {1\over 2} \int_\Lambda {d^3p\over
  (2\pi)^3} \;{1\over p^2} 
\ee
where we introduced the cutoff $\Lambda$ to regulate the divergent integral.
Then the quantity $f(L) - f(0)$ is finite, and can be written in terms of
the renormalized coupling constant $g_R$ via Eqs.~(\ref{eqgap}) and
(\ref{eqRG}).

At a critical point, we have $1/g_R \to 0$. In that case, the gap
equation~(\ref{eqgap}) 
becomes
\be
{1\over 2L} \;\sum_{n=-\infty}^\infty \int {d^2 \bar p\over (2\pi)^2}
{1\over \bar p^2 + \omega_n^2 +\sigma_s} -  {1\over 2} \int_\Lambda
{d^3p\over (2\pi)^3} \;{1\over p^2} = 0 
\ee
which has a non-trivial solution, $\sigma_s = m_*^2$, where
\be
\label{pente}
m_{*}L=2\ln\tau \quad,\quad \tau = \frac{1+\sqrt{5}}{2}
\ee
At this critical point,
the $O(N)$ vector model coupling stays at its
zero temperature critical value $g_*$, (this is sometimes referred to
as the ``finite-size scaling regime'' \cite{Sachdev2}).
One then obtains
the remarkable result~\cite{Sachdev1}
\bea
\label{tesera}
\frac{C(g_{*})}{N} & = & \frac{1}{2}\int
\frac{\rmd^{3}p}{(2\pi)^{3}} \ln p^{2} -\frac{1}{2L}\sum_{n=-\infty}
^{\infty}
\int\frac{\rmd^{2}\bar{p}}{(2\pi)^{2}}\ln(\bar{p}^{2}+\omega_{n}^{2}
+m_{*}^{2})
+\frac{1}{2}\int\frac{\rmd^{3}p}{(2\pi)^{3}}\frac{m_{*}^{3}}{p^{2}}
\nonumber \\
& = & \frac{1}{2\pi L^{3}}\left[{\rm Li}_{3}(e^{-m_{*}L})-\ln(e^{-m_{*}L})
{\rm Li}_{2}(e^{-m_{*}L})+\frac{1}{6}\; (m_{*}L)^3\right] \nonumber \\
& = & \frac{4}{5}\frac{\zeta(3)}{2\pi L^{3}}
\eea
The simplicity of this
result is due to some remarkable identities for
the dilogarithm and 
the trilogarithm at the special point Eq.~(\ref{pente}) \cite{Lewin} [See
Appendix].  

As the free-energy density of a theory of $N$ massless free scalars is
\be
\label{exi}
\frac{C_{0}(g_{0})}{N}=\frac{\zeta(3)}{2\pi }
\ee
one is tempted to conclude from Eqs.~(\ref{tesera}) and (\ref{exi})
that, for large-$N$,  the number of massless degrees of freedom
coupled to the non-trivial fixed-point of the $O(N)$ vector models, is
4/5 times the massless degrees of freedom coupled to the free-field
theory fixed-point. 
However, the strongly coupled critical theory is not related
in an entirely obvious way to the free-field theory of $N$ massless
scalars. It has been argued \cite{Zinn-Justin} that the
critical UV-stable fixed point of the $O(N)$ vector model is in the
same universality class with the
IR-stable fixed point of the $O(N)$ invariant $\phi^{4}$ theory in
$2<d<4$. Such a correspondence is only valid at the critical
regime when one can disregard an infinite number of irrelevant
operators. Therefore, the RG-flow from the free-field
theory of $N$ massless bosons, (corresponding to the UV-stable fixed
point of the $O(N)$ invariant $\phi^{4}$ theory), to the IR-stable
fixed point is non-perturbative. This means essentially that had we
been able to obtain the full solution of the RG-equations, the massless
free theory would correspond to the weak-coupling limit and the non-trivial
critical theory Eq.~(\ref{tesera}) to the strong coupling limit.

\subsection{The $U(N)$ Gross-Neveu Model}

Another three-dimensional theory with a strongly coupled fixed point
is the $U(N)$ Gross-Neveu model \cite{Zinn-Justin,Rosenstein}. Its
Lagrangian is written is terms of 
an auxiliary scalar field $\lambda$ as
\be \label{GN}
{\cal L}=-\bar{\psi}^{\a}(\gamma_{\m}\partial^{\m}+\lambda)\psi^{\a}
+\frac{N}{2G} \lambda^2
\ee
where $\a=1,2,..,N$ \footnote{For the Euclidean three-dimensional
  gamma matrices we use the Hermitian two-dimensional 
  representation  $\gamma_{1}=\sigma^1$,
  $\gamma_{2}=\sigma^2$, 
  $\gamma_{3}\equiv\gamma_{0}=\sigma^{3}$, with $\sigma^{i}$, $i=1,2,3$
  the usual Pauli matrices.} This theory has a UV-stable critical
point at each order in the $1/N$ expansion as the coupling $G$ is kept
fixed. This critical point represents a space-parity breaking transition.
   
Placing the theory in a slab with one finite dimension of length $L$ 
amounts to
imposing antiperiodic boundary conditions for the fermions along 
Euclidean time. Then, when the coupling stays at its bulk critical
value $G_*$ the gap equation gives zero expectation
value for $\lambda$. This is a consequence of the absence of zero modes
for fermions and antiperiodic boundary conditions. It essentially
means that $C(G_*)$ for this model is given by the free-field theory
result \cite{Vitale} and not by a complicated relation involving
polylogarithms such as Eq.~(\ref{tesera}). Therefore, for some time its
was thought that nothing interesting happens in  three-dimensional
fermionic models at the finite-size scaling regime. 

However, it was recently discovered \cite{Marcello,Tassos2} that
certain three-dimensional fermionic models 
do exhibit non-trivial behavior in the finite-size scaling
regime. The theory discussed in \cite{Tassos2} is the $U(N)$
Gross-Neveu model for fixed total fermion number $B$. The fixed fermion
number constraint is introduced into the theory via a delta function
$\delta(\hat{N} -B)$, where
\be
\hat{N}=\int\rmd^2 \bar{x}{}\psi^{\dagger}(\bar{x}){}\psi(\bar{x})\,,
\,\,\,\,\, \bar{x}=(x_{1},x_{2})
\ee
is the fermion number operator. Using an auxiliary scalar field
$\theta(x_{3})$ to impose the delta function constraint, the
action  for this model is written in the slab geometry
\be \label{GNfixed}
{\cal S} =-\int_{slab} \bar{\psi}^{\a}(\gamma_{\m}\partial^{\m}+{\rm
  i}\gamma_3 
\theta +\lambda)\psi^{\a}
+\frac{N}{2G} \int_{slab}\lambda^2+{\rm i}B\int_{slab}\theta
\ee
It was shown in \cite{Tassos2} that for certain values of
$\theta$ the gap equation of the model yields a non-zero expectation
value for $\lambda$ when the coupling stays at the bulk critical
value $G_*$. Consequently, $C(G_*)$ is no longer given by the free
field theory result but by a non-trivial expression involving
polylogarithms as
\bea \label{GNfe}
\frac{C(G_{*})}{NL^3} & = & -\frac{\langle\lambda\rangle^3}{6\pi}
+\frac{1}{\pi L^3} 
\left[\ln\left( e^{-L\langle\lambda\rangle}\right) {\rm Li}_2\left
    ( -e^{-L\langle\lambda\rangle},L\langle\theta\rangle \right)-
{\rm Li}_3\left 
    ( -e^{-L\langle\lambda\rangle},L\langle\theta\rangle
  \right)\right]\nonumber \\
& & +\frac{\langle\theta\rangle}{2\pi L^3}\left[ Cl_2(2\phi)-Cl_2(2\phi
  -2L\langle\theta\rangle)-Cl_2(2L\langle\theta\rangle)\right]\\
\phi & = & \mbox{arctan}\left[\frac{ e^{-L\langle\lambda\rangle}\sin(L
    \langle\theta\rangle)}{1+e^{-L\langle\lambda\rangle}\cos(L
    \langle\theta\rangle)}\right]
\eea
where ${\rm Li}_n(r,\theta)$ is the real part of the polylogarithm
${\rm Li}_n(re^{{\rm i}\theta})$ and $Cl_n(\theta)$ is Clausen's function
\cite{Lewin}. An important observation of \cite{Tassos2} is that for
$L\langle\theta\rangle =\pi$ the fermions acquire periodic boundary
conditions and the theory resembles the bosonic $O(N)$ vector model
discussed previously. In this case the expectation value
of $\lambda$ is given by 
\be
L\langle \lambda\rangle = 2\ln\tau
\ee
and, from Eq.~(\ref{GNfe}), the free-energy is 
\be \label{nfe}
\frac{C(G_*)}{N}=-\frac{8}{5}\frac{\zeta(3)}{2\pi}
\ee
{\em i.e.,} minus twice the free-energy density of the $O(N)$ vector
model. The interpretation of a negative free-energy density may not be
clear when one considers the model above by itself. It is entirely
justifiable, however, when the model is viewed as part of a 
larger theory containing both fermions and bosons as we discuss in the
next Section.

\section{Stationary trajectories and the free-energy density}
\label{sec3}

The models described in Section~\ref{sec2} are typical examples of
strong-coupling 
dynamics in QFT. The low-energy critical theory is
unreachable by a weak coupling expansion around the UV-critical
point and one needs to know the full analytic dependence of
physical quantities on the coupling constant to study  both
regimes. Most importantly, the spectrum of the theory changes in an
uncontrollable way. 

To understand the difference between the massless degrees of freedom
coupled to the different critical regimes of the $O(N)$ model, we shall
exploit the following idea. We shall construct stationary flows
which interpolate between critical theories with different field
contents, one of which is the free-field theory of $N$
massless scalars.
Since the
free-energy density remains invariant along stationary flows, all
such theories will have the same free-energy density Eq.~(\ref{exi}). These
theories may be interpreted as  
interacting theories of massive scalar fields.
Then, the 
condition for stationary free-energy density implies definite
relations between the masses. For a set of special values of the
masses, the resulting theory is seen to be related to the strongly
coupled critical theory Eq.~(\ref{tesera}).

Consider a theory in three dimensions with nine $O(N/4)$ bosonic fields,
$\vec\phi_I$ ($I=1,\dots,9$),
and two $U(N/4)$ fermionic fields
$\vec\psi_J$ ($J=1,\dots,4$).
These fields have masses $m_I$, ($I=1,\dots,9$) and $m_J'$, ($J=1,2$),
respectively.  The
quadratic part of the Lagrangian density reads
\be\label{model}
{\cal L}_0  = \frac{1}{2}\sum_{I=1}^9 \vec\phi_I\cdot
 (-\partial^2 + m_I^2) \vec\phi_I
-\sum_{J=1}^2\bar\psi_J^{\a} (\gamma_\mu \partial^\mu + m_J') \psi_J^{\a}
\ee
where $\a =1,2,\dots,N/4$.
These fields interact,  but we will not
specify the interaction Lagrangian. Furthermore, we may impose a fixed
total fermion 
number constraint as it was discussed in Section~\ref{sec2}. This is quite
natural since the total fermion number constraint is equivalent to the
presence of a background gauge field \cite{Weiss,Pisarski}. Hence, the
theory in Eq.~(\ref{model}) may be viewed as a three-dimensional
massive gauge
theory with bosons and fermions. The full
Lagrangian density is 
\be
\label{epta}
{\cal L} = {\cal L}_0 + V_{int} (\vec\phi_I\;,\; \psi_J^\a)
\ee
where the interaction potential $V_{int}$ is a function of $O(N/4)$
invariants made out of the bosons and fermionic currents.

Each bosonic
field has free-energy in the saddle point approximation 
\be \label{fdensity}
F_b^{(I)} = {N\over 8} \Tr \ln (-\partial^2 + m_I^2)\,,\,\,\,\,\, I=1,\dots,9
\ee
The corresponding expression for a fermionic field is
\be \label{ffdensity}
F_f^{(J)} = - {N\over 4} \Tr \ln (-\partial^2 + (m_J')^2)\,,\,\,\,\,\, J=1,2
\ee
The total free-energy of the system is
\be \label{totalfe}
F = \sum_{I=1}^9 F_b^{(I)} + \sum_{J=1}^2 F_f^{(J)}
\ee
In the slab geometry bosons acquire periodic boundary conditions along
the finite dimension 
$p_{0}=2\pi n/L, n=0,\pm 1 ,\pm 2,\dots$. Furthermore, due to the total
fermion number constraint \cite{Tassos2}, or equivalently due to the some 
special configuration of the background gauge field \cite{Pisarski},
we can arrange so that fermions acquire 
periodic boundary conditions as well. Then using the identity 
\be
\sum_{n=-\infty}^\infty \ln (\omega_n^2 + |\vec p|^2 + m^2) = 2\ln \left(1-
e^{-L\sqrt{|\vec p|^2+m^2}} \right) + L\int {dp_0\over 2\pi}
\ln (p_0^2+|\vec p|^2+m^2)
\ee
we obtain
\be
F_b^{(I)} = {N\over 4L} \int {d^2\vec p\over (2\pi)^2} \ln
\left(1- e^{-L\sqrt{|\vec p|^2+m_I^2}} \right) + {1\over 2}
\int {d^3p\over (2\pi)^3} \ln (p^2+m_I^2)
\ee
and similarly for the fermions.
These expressions are divergent. We will subtract the UV divergence {\em\`a la}
Bogoliubov. To this end, we expand
\be
\ln (p^2+m_I^2) = \ln p^2 + {m_I^2\over p^2} + o(m_I^4)
\ee
and replace
\be
\ln (p^2+m_I^2) \rightarrow \ln (p^2+m_I^2) - \ln p^2 - {m_I^2\over
p^2}
\ee
The resulting expression is finite and given by
\be
8\pi L^3 F_{b\; R}^{(I)}/N = {\rm Li}_3 (e^{-m_IL}) - \ln (e^{-m_IL})
{\rm Li}_2 (e^{-m_IL}) - {1\over 6} (m_IL)^3
\ee
Corrections to this expression may come from the interactive part
of the Lagrangian Eq.~(\ref{epta})
and are in general polynomials in the masses. As
long as we are interested in stationary trajectories, we will
consider only marginal couplings which are trinomials in the 
masses. We will add appropriate trinomials to obtain a free-energy
which is invariant along a certain trajectory in the space of couplings.

Pick a RG trajectory ($\beta (g)=0$) parametrized by $t$
\footnote{This plays now the role of the new parameter $t'$ discussed
  in the Introduction.}.
Along this trajectory, we demand that the function
\be
\label{enia}
C = 2\pi {L^3\over N} \left( \sum_{I=1}^9 F_{b\; R}^{(I)} + \sum_{J=1}^2
F_{f\; R}^{(I)}\right) +
L^3 \sum_{i,j,k} \alpha_{ijk} m_im_jm_k
\ee
be invariant. 
It turns out that a convenient choice is
\be\label{dunamiko}
\sum_{i,j,k} \alpha_{ijk} m_im_jm_k = \sum_{I=1}^4 \left( {1\over 3} m_I^3
-{1\over 2} m_Im_{I+4}^2 - {1\over 2} m_I^2 m_{I+4} + {1\over 3} m_{I+4}^3
\right)
\ee
Defining
\be
x_I = e^{-m_IL} \quad (I=1,\dots,9)\;, \quad\quad
y_J = e^{-m_J'L} \quad (J=1,2)
\ee
and setting
$$
x_1 = x_2 = x_3 = x_4 = t \quad,\quad x_5 = x_6 = x_7 = x_8 = 1-t \quad,\quad
x_9 = {t^2\over (1-t)^2}$$
\be\label{mazes}
y_1 = y_2 = {t\over 1-t}
\ee
we obtain
\be\label{stathero}
{dC\over dt} = 0
\ee
which is the result of non-trivial polylogarithm identities
[see Appendix].

Next, we discuss the behavior of the theory along the trajectory.
The flow of the masses is shown in Fig.~1.
To consider theories with real, non-negative masses, the parameter $t$
must be
restricted to the interval
\be
0 \leq t \leq {1\over 2}
\ee
Along the trajectory, $0<t<1/2$, we have $O(N/4)$ symmetry. At the two ends,
$t=0,1/2$, we obtain enhanced mass degeneracy. There is also a special
point in the interval $(0,1/2)$ where mass degeneracy is enhanced and the
symmetry group becomes $O(5N/4)$.

As $t\to 0$, we have $x_1,x_2, x_3,x_4,x_9\to 0$, as well as
$y_1,y_2\to 0$. Therefore, the bosonic fields
$\vec\phi_1,\vec\phi_2, \vec\phi_3,\vec\phi_4,\vec\phi_9$ and all
four fermionic fields become infinitely massive and decouple.
The other four bosonic fields have $x_I\to 1$ ($I=5,\dots,8$),
so they all become massless. Therefore, in this limit we obtain $4 (N/4) = N$
free massless bosonic fields. Eq.~(\ref{enia}) becomes
\be
C = C(0) = {\rm Li}_3 (1)
\ee
and the
underlying theory which flows to the $t=0$ point of the above
stationary RG-flow is the free theory of $N$ massless bosons.

There is an interesting point in the interval $0 \leq t \leq {1\over 2}$
at which the masses are divided into two sets, all masses in the same set
converging to the same value.
Let $\tau$ be the ``golden mean"
\be
\tau^2 - \tau - 1=0 \quad,\quad \tau = {1+\sqrt 5\over 2}
\ee
At $t = 1/\tau^2 = 2-\tau$, we have
$$
x_1 = x_2 = x_3 = x_4 = x_9 = 2-\tau \quad,\quad x_5 = x_6 = x_7 = x_8
= \tau - 1 $$ 
\be
y_1 = y_2 = {1\over \tau^2 -1} = {1\over \tau}
= \tau -1
\ee

At this point, Eq.~(\ref{enia}) becomes

\be
\label{okto}
C = {5\over 4} \left( {\rm Li}_3 (2-\tau) - \ln (2-\tau){\rm Li}_2
(2-\tau) +{1\over 6} 
\ln^3 (2-\tau) \right)
\ee
However, this is nothing but 5/4 times Eq.~(\ref{tesera}), the latter giving
the free-energy of the UV critical point of the $O(N)$ vector
model. This means that the underlying theory flowing to the $t=2-\tau$
point of the stationary flow above corresponds to 5/4 copies of the $O(N)$
vector model. Consequently, the degrees of freedom coupled to the
non-trivial UV critical point of the $O(N)$ vector model are 4/5 times
the degrees of freedom coupled to the free massless theory of $N$
scalars {\it q.e.d.}

\section{Discussion and Outlook}
\label{sec4}

In this work, we addressed the issue of non-perturbative RG flows
connecting weakly and strongly coupled fixed-points of
three-dimensional QFTs. A quantity which encodes important information for such
fixed points is the free-energy density. We briefly discussed  
some of its properties in Section~\ref{sec2} and   
analyzed some old \cite{Sachdev1}, and more recent \cite{Tassos2} exact
results for the free-energy density of non-trivial, strongly coupled
three-dimensional CFTs in Section~\ref{sec3}. We presented our main result in
Section~\ref{sec4}, where we provided an explicit example of a
stationary trajectory 
interpolating between a
class of critical three-dimensional QFTs. The crucial observation is that
the interactive critical theory corresponding to a special
point of the trajectory  is related to the non-trivial critical point
of the three-dimensional $O(N)$ vector model. This way, we provided a
physical picture for the fact that the latter critical theory seems
to involve 4/5 times the number of
massless degrees of freedom of the free-field
theory of $N$ massless scalars. Our arguments are based on the
observation that the stationary points of the quantity $C(t)$
correspond to possible critical points. The introduction of the finite
length $L$ was also essential in order to obtain more than one
non-trivial stationary points for $C(t)$.

Along the stationary flow, we have a $O(N/4)$
symmetry. At the two ends of the flow (see Fig.~1), we
have enhanced mass
degeneracy. At one end, we obtain $4(N/4)=N$ free massless bosons; at the other
end, we obtain an enhanced symmetry, $O(5N/4)$ and the model rolls to a
non-trivial conformal field theory. In obtaining the stationary trajectory of
Eqs.~(\ref{enia}) and (\ref{stathero}) it was essential that the
fermions in the Lagrangian Eq.~(\ref{model}) acquire periodic boundary
conditions. This is not unusual for
fermions in gauge field backgrounds, as the zeroth component of the
gauge field plays the role of an imaginary chemical
potential. Therefore, it is possible that for certain configurations of
the background gauge field the fermions acquire a zero mode and 
are effectively bozonized contributing, however, a negative free-energy
density  \cite{Tassos2,Pisarski}. Although the issue regarding the physical
interpretation of such a bosonization
is not clearly settled as yet \cite{Smilga}, we
have explicitly demonstrated
its relevance to studies of non-perturbative RG flows.
  
Our approach to studying the free-energy density of
critical theories connected via non-perturbative RG-flows
could in principle be applied to other interesting cases, such as the
${\cal{N}}=4$ SYM$_{4}$, in connection with the adS/CFT correspondence.
There, it is found that the free-energy
density of the ${\cal N}=4$ SYM is $3/4\times$ the free massless
theory. This is a puzzle with no apparent resolution. Our results suggest
that in order to address this issue, one might try to construct
a stationary flow, within the
massless-free theory and see whether at some point it becomes
equal to $4/3\times$ the free energy of a non-trivial CFT. Such an extension 
of our results
is expected to be more complicated, as the corresponding formulae for
the free-energy density of massive four-dimensional theories involve
infinite sums of polylogarithms \cite{Actor}. A further complication
comes from the fact that the beta-function of ${\cal N}=4$ SYM$_{4}$
is identically zero. Therefore, neither the 
RG Eq.~(\ref{duo}), nor any kind of weak-coupling expansion  are
sufficient to studying the coupling constant 
dependence of $C(g,l)$. In this respect, all efforts towards
understanding the non-perturbative nature of ${\cal N}=4$ SYM$_4$ are
worthwhile.

\newpage
\appendix   
\section{Some Polylogarithmic Identities}
   
Here we present certain polylogarithm identities which
we used in order to derive various results earlier.
For more details see~\cite{Lewin}.

The $n$th-order polylogarithm is defined by
\be
{\rm Li}_n (x) = \sum_{k=1}^n {x^k\over k^n}
\ee
We have ${\rm Li}_n (1) = \zeta (n)$.

The dilogarithm obeys the following identities:
\be
\label{app1}
{\rm Li}_2 (x)+ {\rm Li}_2 (1-x) = {\rm Li}_2 (1)
-\ln x \ln (1-x)
\ee
\be
\label{app2}
{\rm Li}_2 (x)+ {\rm Li}_2 (-x) = {1\over 2}
{\rm Li}_2 (x^2)
\ee
\be
\label{app3}
{\rm Li}_2 (x) + {\rm Li}_2 \left( {-x\over 1-x}\right)
= - {1\over 2} \ln^2 (1-x)
\ee
The golden mean plays a special role in these identities.
Let $\tau = {1+\sqrt 5\over 2}$. Setting $x= 1/\tau$
in the first two identities (Eqs.~(\ref{app1}) and (\ref{app2})), we obtain
\be
\label{app4}
{3\over 2} {\rm Li}_2 (1/\tau^2) - {\rm Li}_2 (-1/\tau)
= {\rm Li}_2 (1) - {1\over 2} \ln\tau^2
\ee
where we used $\tau^2 - \tau - 1 =0$.
Setting $x= 1/\tau^2$ in the third identity (Eq.~(\ref{app3})), we obtain
\be
\label{app5}
{\rm Li}_2 (1/\tau^2) - {\rm Li}_2 (-1/\tau) = - {1\over 8}
\ln^2 \tau^2
\ee
Combining Eqs.~(\ref{app4}) and (\ref{app5}), we finally
obtain
\be
\label{app8}
{\rm Li}_2 (1/\tau^2) = {2\over 5} {\rm Li}_2 (1) - {1\over 4}
\ln^2\tau^2
\ee
{\em i.e.}, the dilogarithm can be expressed in terms of
elementary functions at the special point $x=1/\tau^2$.

Similarly, for the trilogarithm, the following identities
hold:
\be
\label{app6}
{\rm Li}_3 (x)+ {\rm Li}_3 (1-x) + {\rm Li}_3 \left( {-x\over
    1-x}\right) = {\rm Li}_3 (1) 
+ {\rm Li}_2 (1) \ln (1-x)
-{1\over 2} \ln x \ln^2 (1-x) + {1\over 6} \ln^3 (1-x)
\ee
\be
\label{app7}
{\rm Li}_3 (x)+ {\rm Li}_3 (-x) = {1\over 4}
{\rm Li}_3 (x^2)
\ee
Again, the golden mean plays a special role. To see this, set
$x = 1/\tau^2$ in the first identity (Eq.~(\ref{app6}))
\be
{\rm Li}_3 (1/\tau^2)+ {\rm Li}_3 (1/\tau) + {\rm Li}_3 (-1/\tau) =
{\rm Li}_3 (1) 
+ {1\over 2} {\rm Li}_2 (1) \ln \tau^2
- {5\over 48} \ln^3 \tau^2
\ee
Using the second identity (Eq.~(\ref{app7})), we obtain
\be
{\rm Li}_3 (1/\tau^2) = {4\over 5} {\rm Li}_3 (1) - {2\over 5}
{\rm Li}_2 (1) \ln \tau^2 - {1\over 12} \ln^3 \tau^2
\ee
which, in view of Eq.~(\ref{app8}), can be written as
\be
\label{app9}
{\rm Li}_3 (1/\tau^2) + {\rm Li}_2 (1/\tau^2) \ln \tau^2 - {1\over 6}
\ln^3 \tau^2 
= {4\over 5} {\rm Li}_3 (1)
\ee
Introducing the function
\be
{\cal F} (x) = {\rm Li}_3 (x) - {\rm Li}_2 (x) \ln x + {1\over 6} \ln^3 x
\ee
we can write Eq.~(\ref{app9}) as ({\em cf.}~Eq.~(\ref{tesera}))
\be
{\cal F} (1/\tau^2) = {4\over 5} \; {\cal F} (1) = {4\over 5} \; \zeta (3)
\ee
To show that the function Eq.~(\ref{enia}) is invariant under the flow
Eq.~(\ref{mazes}), set

$x=t$ in Eq.~(\ref{app6}) and $x= {t\over 1-t}$ in
Eq.~(\ref{app7}). Combining the 
two equations, we obtain
$$
{\rm Li}_3 (t) + {\rm Li}_3 (1-t) + {1\over 4} {\rm Li}_3 (t^2/(1-t)^2) -
{\rm Li}_3 (t/(1-t)) $$
\be
= {\rm Li}_3 (1) + {\rm Li}_2 (1) \ln (1-t) - {1\over 2} \ln t
\ln^2 (1-t) + {1\over 6} \ln^3 (1-t)
\ee
Now define the symmetric function ({\em cf.}~Eq.~(\ref{dunamiko}))
\be
{\cal V} (x,y) = {1\over 3} \ln^3 x - {1\over 2} \ln^2 x \ln y
- {1\over 2} \ln x \ln^2 y + {1\over 3} \ln^3 y
\ee
After some algebra, we obtain
\be\label{telos}
{\cal C} = {1\over 4} \sum_{I=1}^9 {\cal F} (x_I) -
{1\over 2} \sum_{J=1}^2 {\cal F} (y_J) - \sum_{I=1}^4 {\cal V}
(x_I,x_{I+4}) = {\rm Li}_3 (1) 
\ee
where the $x_i$ ($i=1,\dots,9$), $y_1,y_2$ are given by Eq.~(\ref{mazes}).
Eq.~(\ref{telos}) implies stationary flow ({\em cf.}~Eq.~(\ref{stathero})),
\be
{d {\cal C}\over dt} = 0.
\ee


\end{document}